\documentclass[reprint]{aastex6}
\usepackage{natbib}
\slugcomment{Accepted for publication in  The Astrophysical Journal}
\shorttitle{TRANSIT POLARIZATION OF TRAPPIST-1}
\shortauthors{SENGUPTA}
\received{10 February, 2018}
\accepted{16 May, 2018}
\begin{document}

\title{POLARIZATION OF TRAPPIST-1 BY THE TRANSIT OF ITS PLANETS}

\author{Sujan Sengupta}
\affil{Indian Institute of Astrophysics, Koramangala 2nd Block,
Bangalore 560 034, India; sujan@iiap.res.in}

\begin{abstract}
 
As the first and till date the only one multiple planet hosting  dwarf star that is 
sufficiently cool to form  condensate cloud in it atmosphere, Trappist-1 provides unique
opportunity to test the efficiency of image polarimetry as a tool to detect and
characterize exoplanets around L- and late M-dwarfs and Exomoons around directly imaged
self-luminous giant exoplanets. Although scattering of light by atmospheric dust
particles should produce significant amount of linear polarization in the far optical
and near infra-red, the disk-averaged net detectable polarization of the star must be
zero owing to spherical symmetry. However, the transit of its planets would give rise
to significant asymmetry and produce phase-dependent polarization with the peak 
polarization occurring at the inner contact points of planetary transit ingress and egress epoch. 
Adopting the known stellar and planetary physical parameters and employing
a self-consistent cloudy atmosphere model of M8 dwarf star, the transit  
polarization profiles and the expected amount of polarization of Trappist-1
during the transit phase of each individual planets as well as that during simultaneous transit
of two planets are presented in this paper. It is emphasized
that the amount of polarization expected is within the detection limit  of a few existing
facilities. If polarization is detected confirming the prediction, time resolved image
polarimetry will emerge out as a potential tool to detect and characterize small planets
around cloudy ultra-cool dwarfs.

\end{abstract}

\keywords{stars: individual: TRAPPIST-1 --- stars: low-mass --- polarization
--- scattering --- planets and satellites: detection  --- occultations}

\section{INTRODUCTION}

Two decades after the first confirmed discovery of planets outside the solar system
\citep{wolszczan92,wolszczan94,mayor95}, more than 3000 planets of different size,
mass and composition are detected. Most of these planets are found to be orbiting around
solar-type star.
It is known that Red dwarfs and Brown Dwarfs account for about 80 \% of the 
stellar population in our galaxy \citep{apai13}.
As a consequence search for planets around these ultra-cool dwarfs will enable 
to test the formation models because it'll provide a complete and comprehensive knowledge  on the 
planetary architecture as a function of the mass of the host.
Therefore, the discovery of 7 Earth-sized planets transiting the M8V
star Trappist-1 \citep{Gillon16, Gillon17} is a significant step. Trappist-1 
(2MASS J23062928-0502285) is located 12.14 pc from the Sun \citep{grootel18}.
It is the first planetary system found to transit
a very  low mass, Jupiter-sized star. The favorable planet-to-star ratio offers the opportunity 
to probe their atmospheric properties with the
current and next-generation telescopes \citep{dewit2016,Barstow2016}.
Furthermore, a unique system  such as Trappist-1 enables to  test planet formation and evolution
theories and to investigate the prospects for habitability among Earth-sized exoplanets orbiting cool
M- dwarfs.

The atmospheres of dwarf stars with spectral type M7 and later are cold enough for the condensation
of silicate cloud \citep{tsuji96, alard01, ack01}.
The spectra of L-dwarfs are well explained by the inclusion of condensate cloud of various
species such as iron, forsterite, and enstatite \citep{cushing08,marley15}.
Scattering of thermal radiation by these
dust particles in the visible atmosphere of ultra-cool dwarfs gives rise to significant amount
of linear polarization in the far-optical and in the infra-red regions.
Linear polarization has been detected in the
optical bands from a good number of L-dwarfs covering almost the entire range of spectral
types L0--L8 \citep{men02,oso05,tata09,oso11,miles13}. Scattering by horizontally
homogeneous clouds in the atmosphere of a rotation-induced oblate L-dwarf can explain the
observed polarization \citep{sengupta01,sengupta03,sengupta05,sengupta10}.  

  Similar to the L-dwarfs, Trappist-1, a dwarf star of M8V spectral type 
with effective temperature about 2500 K and about the size of Brown Dwarfs, is expected to
have thin cloud in the upper atmosphere. Therefore, the thermal radiation of Trappist-1
should also be linearly polarized due to scattering by dust particles. However, as a slow
rotator with a spin rotation period as high as 79.2 h \citep{luger17}, the object
should be almost spherical and hence the disk-integrated polarization
of Trappist-1 should be negligible. But, transit by the planets would give rise to
significant asymmetry by blocking the stellar disk partially during the transit epoch 
\citep{sengupta16a}. As a consequence, transit of planets around Trappist-1 would  give 
rise to net non-zero time-dependent disk integrated polarization. Since Trappist-1 is
presently the only discovered ultra-cool dwarf with a cloudy atmosphere and with a system of
seven rocky planets, it can very well serve as a test object for verifying the potential
of image polarimetry as a tool to detect planets around similar late M-dwarfs and L-dwarfs
as well as Exomoon around self-luminous giant exoplanets.

In this paper, I present the transit polarization profiles for Trappist-1 with specific 
prediction of the expected amount of polarization. In the next section, the 
employed atmosphere model and the adopted formalism for calculating the transit polarization
is described. In section~3, I discuss the results and in section~4, I conclude that if confirmed,
time-resolved image polarimetry will open up a new avenue to detect and characterize exoplanets
around ultra-col dwarfs.  

\section{METHOD FOR CALCULATING THE TRANSIT POLARIZATION}

The disk-integrated polarization of a spherical star is zero. However, when a planet
transit the star, the net polarization  during the transit is  non-zero owing to the asymmetry caused
by partial blocking of the stellar disk. The transit
polarization of the star is the product of 
the intensity and scattering polarization at each radial point along the disk of the star 
with  the fractional circumference occulted by
the projection of the planet over the surface of the star.  
The disk-integrated polarization during the transit phase 
is given by \cite{sengupta16, sengupta16a,wik14,car05} : 
\begin{eqnarray}
p_(t,\lambda) =\frac{1}{F}\int^{r_2}_{r_1}2\sqrt{\frac{[(1-\mu^2)^{1/2}-r_m(t)]^2-w^2}{1-\mu^2}}
I(\mu,\lambda)p(\mu,\lambda)\mu d\mu,
\end{eqnarray}
where  $t$ is the time since mid-transit and $F$ is the flux of the unobscured star at a 
given wavelength or band. In the above expression,
$I(\mu,\lambda)$ and $p(\mu,\lambda)$ are the wavelength dependent specific intensity 
and polarization respectively along $\mu$,
$\mu=\cos\theta=\sqrt{1-r^2}$ with $\theta$ being 
the angle between the normal to the surface of the star  and the line of sight and
$r$ being the radial points along the disk of the primary, $0\leq r \leq 1$,     
$r_1=\sqrt{1-[r_m(t)+w]^2}$ and $r_2=\sqrt{1-[r_m(t)-w]^2}$,
$r_m(t)$ is the instantaneous position of the center of the planet and is given by
\begin{eqnarray}
r_m(t)=\left[b^2+4\left\{(1-2w)^2-b^2\right\}\left(\frac{t}{\tau}\right)^2
\right]^{1/2},
\end{eqnarray}
where $b=a\cos i/R_\star$ is the impact parameter of the planet with a circular orbit of
radius $a$ and the orbital inclination angle $i$
and $R_\star$ is the radius of the star, $w=R_P/R_\star$ is the ratio of the planetary 
radius ($R_P$) to the radius of the star and $\tau$ is the total transit duration 
given by \cite{scharf09}
\begin{eqnarray}
\tau=\frac{P}{\pi}\sin^{-1}\left[\frac{R_p}{a}\left\{\frac{(1+w)^2-b^2}{1-
\cos^2 i}\right\}^{1/2}\right],
\end{eqnarray}
$P$ being the orbital period of the planet.
The physical parameters used are presented in table~1. These are derived by using the
values of $R_P$, $a$, $P$ and $i$ as given in \cite{Gillon17}. 

In order to calculate the specific intensity $I(\mu,\lambda)$ and scattering polarization
$p(\mu,\lambda)$ for Trappist 1,
I have employed the well-used
 one-dimensional, non-grey, hydrostatic and radiative-convective
 atmospheric model for the relatively hotter L-dwarfs developed by \cite{ack01,marley02,freed08}.  
This model fits reasonably well
the spectra and photometry of a large number of cloudy L  dwarfs at a wide range of
wavelengths covering near optical to mid-infrared regions \citep{marley15}.
The effective temperature of Trappist-1 is $T_{\rm eff}= 2516\pm 41$ K 
and surface gravity $g=1686.55 {\rm ms^{-2}}$ \citep{grootel18}. I adopt
a grid of atmosphere model with $T_{\rm eff}=2400$ K and $g=1000 {\rm ms^{-2}}$. The 
slightly lower temperature compensates the cloud thickness for a lower surface gravity
and hence the model atmosphere does not differ significantly from that of the actual
atmosphere of Trappist-1. In fact it is known that the synthetic spectra of L-dwarfs or
ultra-cool dwarfs such as Trappist-1 (M8) remain the same for a wide range of surface
gravity. Figure~1 (upper panel) presents a comparision of the theoretical spectrum to the
observed spectral energy distribution of Trappist-1 \citep{bur15}. Owing to the low spectral
resolution of the thoeretical spectrum, a better fit to the observed spectrum at wavelengths
longer than 1.4 micron is not achieved. However, polarization changes slowly with wavelength
and this resolution is adequate for the present purpose. Note that the polarization profiles
are presented in I- and J-band by using the response functions of SDSS filters. The I- and 
J-bands cover wavelength region from 0.65 to 1.4 micron. Therefore, the models
serve the present purpose well.   

In this atmosphere model, spatially uniform dust cloud is included self consistently.
The efficiency of sedimentation of cloud particles in the atmospheric models
is controlled through a scaling factor $f_{\rm sed}$.
In the present work I adopt $f_{\rm sed}=2.0$ \citep{cushing08,stephens09} that 
incorporates moderately thick condensate cloud. However, in order to investigate how 
sensitive the polarization is to the cloud opacity, I have also calculated polarization
with 5\%, 10\% and 15\%  increase in the dust opacity without changing gas opacity in
the model. This does not affect the emergent spectra of the star as shown in Figure~1
(bottom panel) but alters the amount of polarization significantly.

The detailed formalisms as well as the numerical methods for calculating the angle 
dependent intensity and linear polarization are described  in 
\cite{sengupta09, sengupta10}.

\section{RESULTS AND DISCUSSIONS}

The formation of dust grains in the cloudy atmosphere of ultra-cool
dwarfs provides an additional scattering opacity to the gas opacity.
The thermal radiation of such cloudy dwarf stars and self-luminous giant exoplanets
is polarized by dust scattering in the far-optical and infra-red wavelengths 
\citep{sengupta10,marley11}. The net non-zero disk integrated polarization of such cloudy objects
arises only during the transit epoch unless the object is non-spherical. However, slow spin rotation
ensures Trappist-1 has almost perfect spherical symmetry and hence the disk integrated polarization
should be zero during non-transit epoch.

 As depicted in Figure~2 and Figure~3, the polarization increases
significantly with the increase in dust opacity even by 5\%  of the original cloud model.
The scattering  opacity is determined by a balance
between the upward turbulent diffusion and downward transport by sedimentation of
condensates and gas. These are governed by the surface gravity and effective temperature
and hence the polarization varies with different
effective temperature and surface gravity.
While polarization reduces with the decrease in surface gravity,
a decrease in temperature causes the cloud base to shift downward yielding a larger  column
of dust grains in the observed atmosphere and hence the polarization increases with the
decrease in effective temperature \citep{sengupta16a}. 
Therefore, while the adopted effective temperature of Trappist-1
is slightly less than the actual value, a reduced value of surface gravity considered should 
yield into the expected amount of polarization of Trappist-1. However, the actual cloud opacity
is highly model dependent and so the polarization profile is calculated with varying  scattering
opacity. The peak polarization at the inner contact points of transit ingress/egress phase
is thus determined by the scattering opacity as well as the size of the transiting planet. The
amount of polarization is also wavelength dependent. Polarization is found to be maximum at J-band.       

As discussed in \cite{sengupta16, sengupta16a}, 
the general feature of transit polarization of any star is
the double peaked polarization profile that arises because 
the maximum polarization occurs at the inner contact points of transit ingress/egress phases. For
central transit i.e., when the impact parameter $b=0$, the projected position of
the center of the planet coincides with the center of the star during mid transit
giving rise to a symmetry to the projected stellar disk. Hence
the disk integrated polarization for central transit or transit with a small value of the impact
parameter $b$ is zero or negligibly small during mid transit. Although, the orbital inclination
angle of all the seven planets are almost the same, owing to a closer orbit, the values of
the impact parameters of Trappist-1b, Trappist-1c, Trappist-1d and Trappist-1e are quite small 
(see table~1) and hence the amount of
polarization during mid transit of these four planets is negligibly small as shown in
Figure~2. However, as the
orbital distance $a$ increases, the impact parameter $b$ increases for $i\neq 0$ 
making the transit significantly off center. Therefore, the polarization is non-zero
during the whole transit epoch including the mid transit time of Trappist-1f, Trappist-1g
and Trappist-1h (see Figure~3).
  
 The polarization increases sharply from zero at the outer contact point 
and peaks at the inner contact point of transit ingress. Similarly, the polarization
falls rapidly to zero at the outer contact point from its peak value at the inner contact
point of transit egress. Therefore, the polarization profile can provide the ingress/egress
duration $\tau_{ingress/egress}$. 
 The asymmetry to the stellar disk decreases as the  planet moves to the center ($t=0$)
from the limb (t=$\pm\tau/2$) of the stellar disk. Hence the disk-integrated polarization decreases.
However, the peak polarization at the inner contact points of transit ingress/egress phases
is independent of the impact parameter but depends on the ratio of the planetary to stellar radii.

  The repeated occurrence of non-zero polarization with double peak would provide the orbital 
period of the planets. The total transit duration can be found from the transit polarization
profile and the impact parameter $b$ for a given radius of the star and orbital distance
of the planet can be determined from  the polarization during the mid transit at $t=0$. 
Hence, the duration of the transit ingress/egress phase will provide an estimation of the size of
the planet independent of any limb-darkening model by using the following expression :

\begin{eqnarray}
\frac{\tau_{ingress/egress}}{\tau_{total}}=\frac{1}{2}-\sqrt{\frac{1}{4}-\frac{w}{(1+w)^2-b^2}}.
\end{eqnarray}

Such a way of estimating the physical parameters of the transiting planet will serve as a 
complementary of and check for the transit method.  
 Further, any stellar activity such as flare
may cause difficulties in confirming the detection of a planet through transit method. But such
activities would not affect the polarization profile in the infra-red. Hence time resolved image
polarimetry may be advantageous over transit method in detecting planets around cloudy ultra-cool
dwarfs.    

  Finally, Trappist-1 has a compact planetary system with several planets orbiting  close
to the star. Therefore, occasionally two or three planets transit the star simultaneousely 
\citep{Gillon16}. Simultaneous transit alters the symmetry and hence the polarization profile. 
In Figure~4 the polarization profile is presented for the case when the planet Trappist-1b
attains its mid transit phase while the planet Trappist-1c touches the inner contact point
of its ingress phase. During the mid transit of Trappist-1b, the symmetry would otherwise give rise
to zero polarization as shown in Figure~2 but the transit of Trappist-1c would provide the
necessary asymmetry and hence the polarzation peaks even during the mid transit (t=0) of
Trappis-1b. Therefore, a triple peak polarzation profile arises in such scenario.
Similarly, I consider another extreme case when one planet reaches the inner contact point of
its egress phase while the second one reaches the inner contact point of its ingress phase. 
Under such situation, the polarization at the inner contact point of the transit egress epoch
of one planet would be the sum of the peak polarization during the
inner contact point of the transit ingress/egress epoch of individual planets as shown in
Figure~5. Although such
occasional event would favour polarimetric observation and detection, repeat image polarimetry
during the transit phase of individual planet is required to estimate the physical properties
of the planet.

\section{CONCLUSIONS}

In this letter, I have presented the transit polarization profiles of Trappist-1 during the
individual transit of all the seven planets around it   
and estimated the expected peak polarization at the inner contact points of transit 
ingress/egress phase for each planet. Polarization profiles during simultaneous transits
of two planets are also presented. For this purpose, the atmospheric
model appropriate for Trappist-1 is employed. The theoretical spectrum calculated by using
the model atomoshere is copmared with the observed spectral energy distribution of Trappist-1
at the optical. I- and J-bands linear polarization due to scattering by dust grains in the
atmosphere of Trappist-1 is calculated by solving the multiple-scattering vector radiative
transfer equations. The model estimation predicts 
detectable amount of transit polarization in both I- and J-bands of
Trappist-1 by the transit of all the planets orbiting it.

Usually it is difficult to detect exoplanets around ultra-cool dwarfs by using the
techniques that are employed to look for exoplanets around solar-like
stars, e.g., radial velocities, transits, direct imaging, or micro-lensing.
Thus, the confirmation of the polarimetric modulation depicted in  Figure~2 and Figure~3 will
open up a new and potential avenue for the search and characterization of exoplanets
around ultra-cool dwarfs. Trappist-1 is currently the only target with the right
properties to investigate the prediction. The predicted amount of polarization can
easily be detected by several existing facilities including FORS2 onboard VLT and LIRIS onboard WHT.

\section{Acknowledgements} 

I thank Mark S. Marley for kindly providing the atmosphere model data and Adam Burgasser for
kindly providing the observed spectrum of Trappist-1. Thanks are due to Enric Palle for many
useful discussions and to the reviewer for providing critical comments and valuable suggestions.

\clearpage
\begin{figure}
\includegraphics[angle=0.0,scale=0.8]{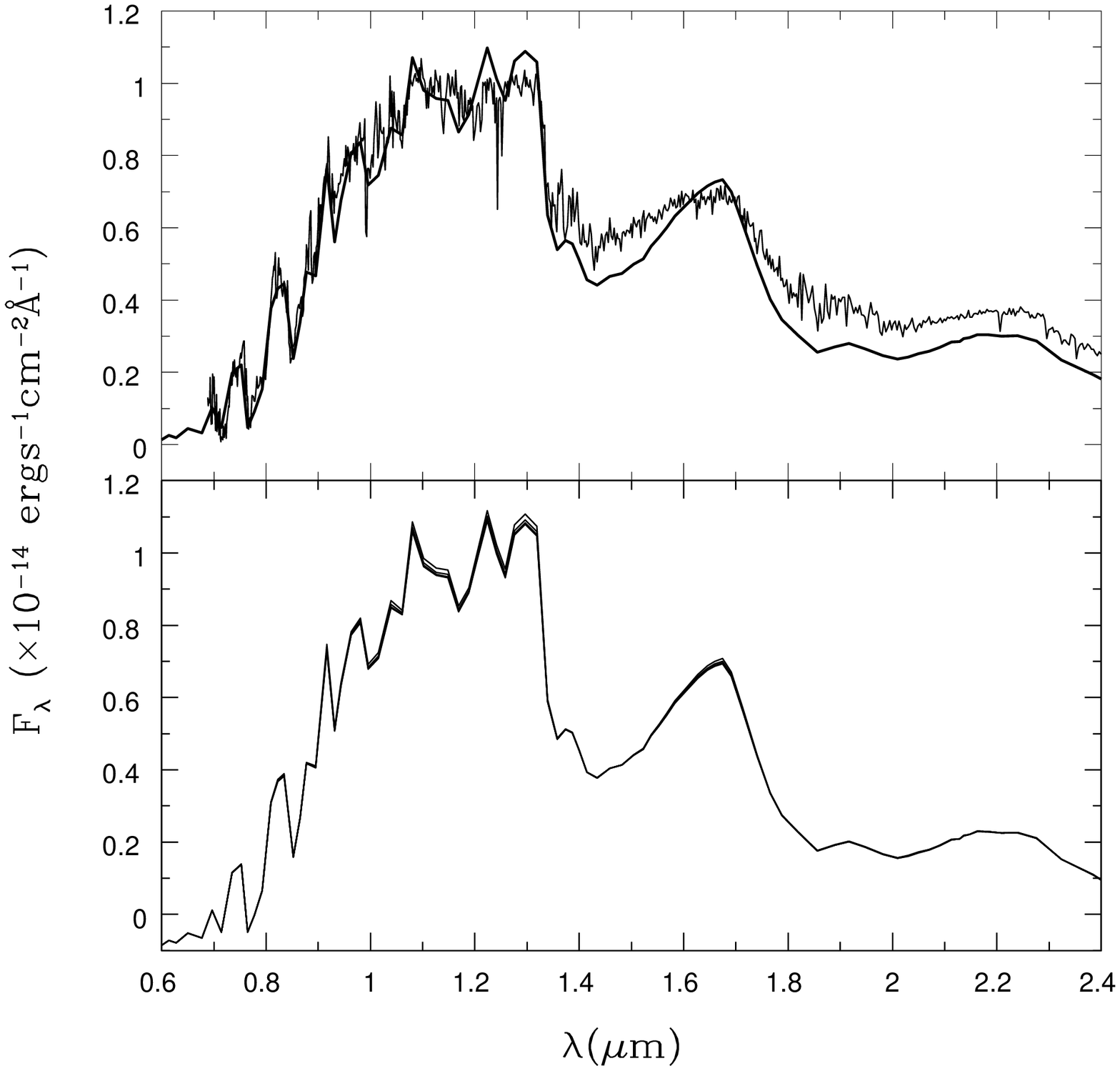}
\caption{Top panel: Comparison of the model spectrum to the observed spectrum of Trappist-1. Thick
line is the theoretical model spectrum and thin line is the observed spectral energy distribution
\citep{bur15}.
Bottom panel : Comparison of the original model spectrum  to that with 5\%, 
10\% and 15\%  increase in the dust opacity.
\label{fig1}}
\end{figure}

\begin{figure}
\includegraphics[angle=0.0,scale=0.8]{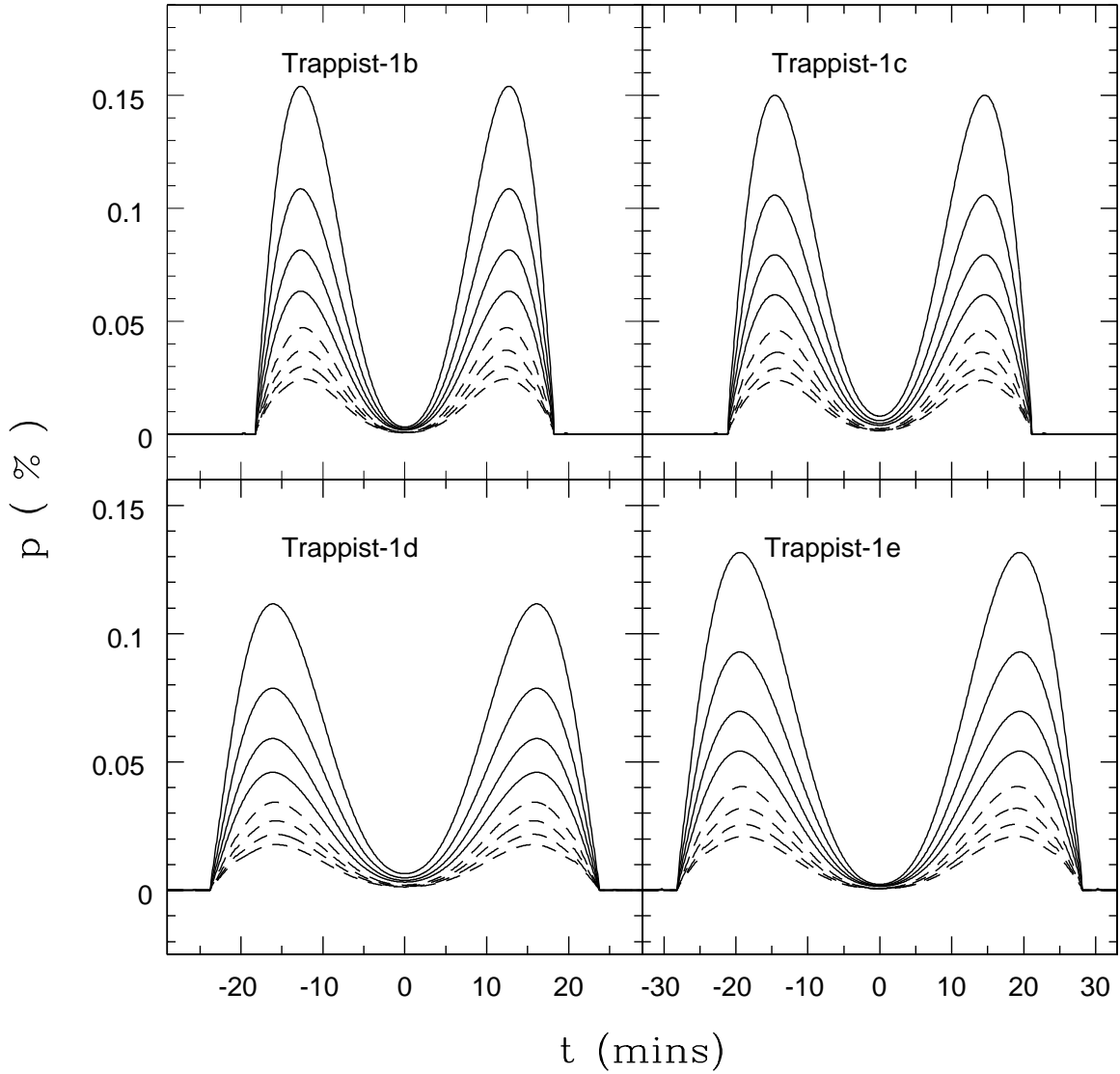}
\caption{I- and J-band polarization of the star Trappist-1 due to the transit of its planets
Trappist-1b, Trappist-1c, Trappist-1d and Trappist-1e. The solid lines represent the
polarization profile in J-band while the broken
lines represent that in I-band. From bottom to top the solid (and broken) lines show the
polarization with the (1) original atmosphere model with $f=2$, (2) 5\% , (3) 10 \%  and (4)
15\%  increased dust opacity.  
\label{fig2}}
\end{figure}

\begin{figure}
\includegraphics[angle=0.0,scale=0.8]{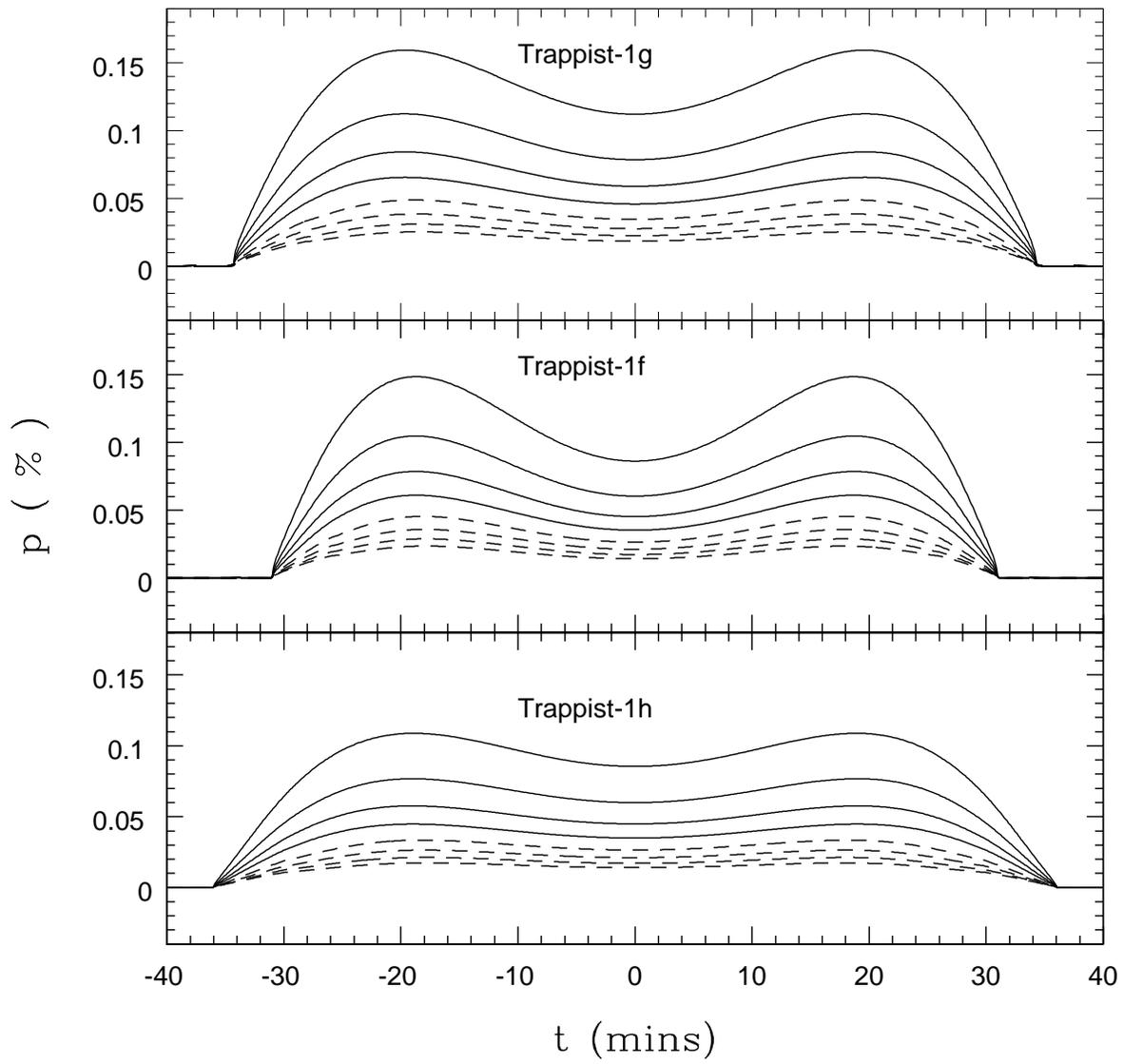}
\caption{Same as figure 1 but due to the transit of planets Trappist-1f, Trappist-1g and
Trappist-1h.
\label{fig3}}
\end{figure}

\begin{figure}
\includegraphics[angle=0.0,scale=0.8]{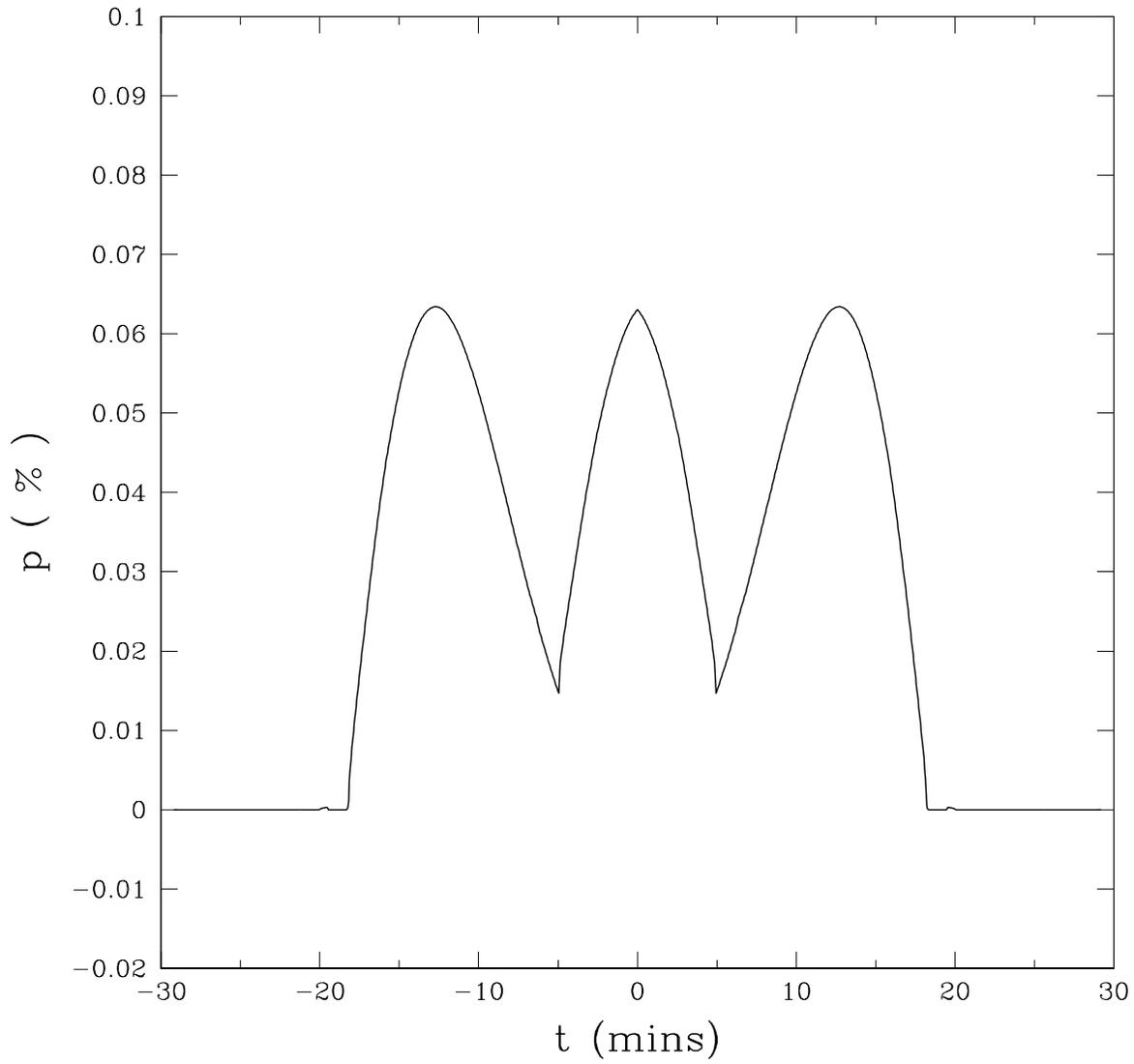}
\caption{Polarization profile during simultaneous transits of Trappis-1b and Trappist-1c. 
Trappist-1b is at mid transit epoch (t=0) when Trappist-1c is at the inner contact point of
the transit ingress epoch.
\label{fig4}}
\end{figure}

\begin{figure}
\includegraphics[angle=0.0,scale=0.8]{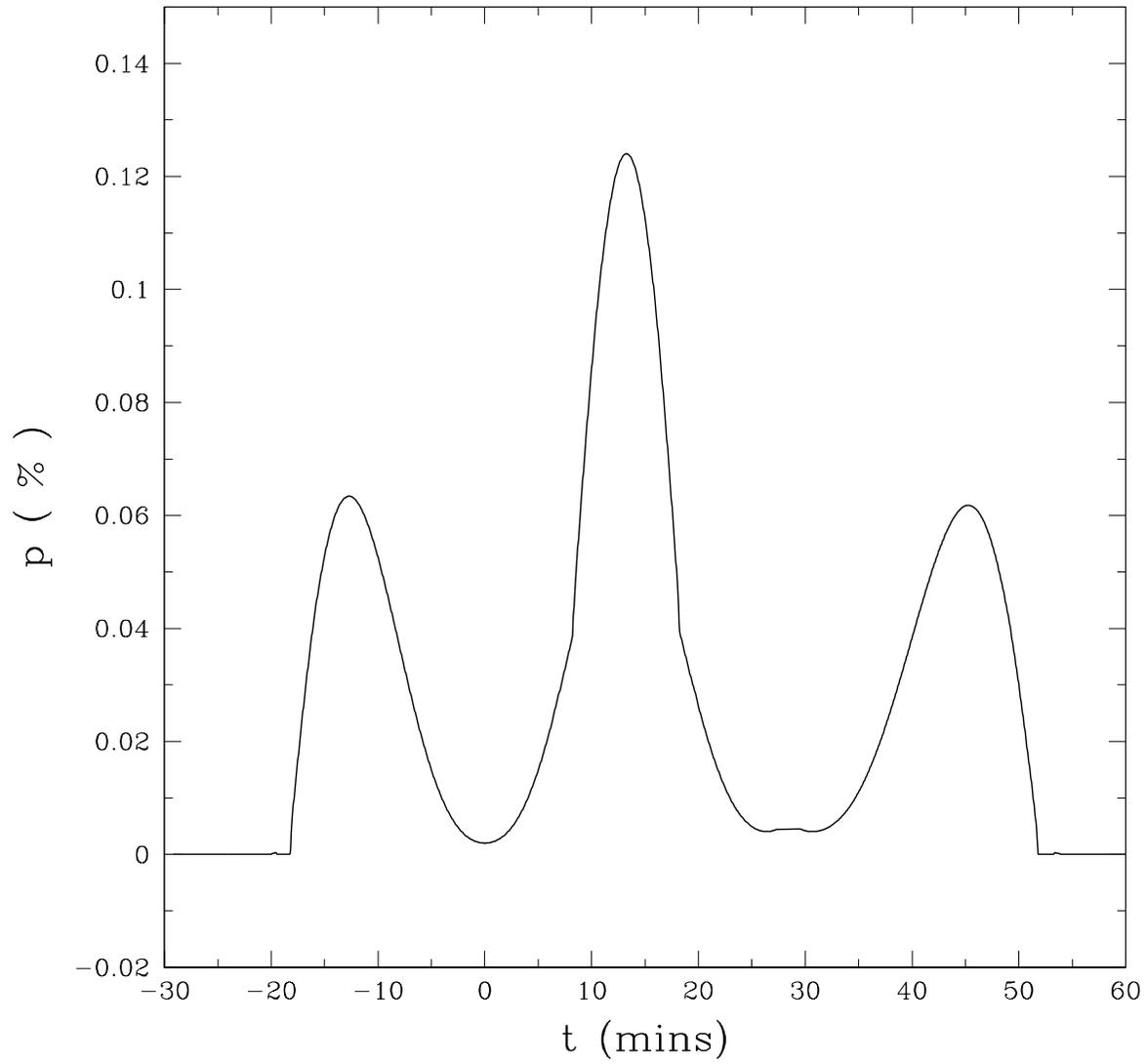}
\caption{Polarization profile during simultaneous transits of Trappis-1b and Trappist-1c. 
Trappist-1b is at the inner contact point of its egress epoch while Trappist-1c is at the
inner contact point of its ingress epoch. t=0 is the mid transit epoch of Trappist-1b.
\label{fig5}}
\end{figure}
\clearpage

\begin{table}
\begin{center}
\caption{Physical parameters of the seven planets transiting Trappist-1}
\begin{tabular}{cccc}
\tableline\tableline
Planets & $w=R_P/R_\star$ &  $b=a\cos i/R_\star$  &  $\tau$ (mins) \\
\tableline
Trappist-1b & 0.085  & 0.125  & 36.48    \\
Trappist-1c & 0.083  & 0,161  & 42.42    \\
Trappist-1d & 0.060  & 0.172  & 49.18    \\
Trappist-1e & 0.072  & 0.126  & 57.34    \\
Trappist-1f & 0.082  & 0.380  & 62.52    \\
Trappist-1g & 0.088  & 0.419  & 68.425    \\
Trappist-1h & 0.059  & 0.454  & 75.697    \\
\tableline
\end{tabular}
\end{center}
\end{table}

\end{document}